\documentclass[reprint,superscriptaddress]{revtex4-1}

\usepackage{graphicx}
\usepackage{amssymb}
\usepackage{amsmath}
\usepackage{epsfig}
\usepackage{chemarr}
\usepackage{url}
\usepackage{natbib}
\usepackage{dcolumn}
\usepackage{bm}
\usepackage{color}

\graphicspath{{GoodFigs_PRXProof/}}

\def\beq{\begin{equation}}
\def\eeq{\end{equation}}

\def\beqa{\begin{eqnarray}}
\def\eeqa{\end{eqnarray}}

\begin{document}
\title{Discriminatory proofreading regimes in non-equilibrium systems}
\author{Arvind Murugan}
\affiliation{Simons Center for Systems Biology, School of Natural Sciences, Institute for Advanced Study, Princeton, NJ 08540, U.S.A.,} 
\affiliation{SEAS, Harvard University, Cambridge, Massachusetts 02138, USA.} 
\author{David A Huse}
\affiliation{Physics Department, Princeton University, Princeton, NJ 08544, U.S.A.} 
\author{Stanislas Leibler}
\affiliation{Simons Center for Systems Biology, School of Natural Sciences, Institute for Advanced Study, Princeton, NJ 08540, U.S.A.,} 
\affiliation{Laboratory of Living Matter, The Rockefeller University, 1230 York Ave., New York, NY 10065, U.S.A.}

\begin{abstract}
We use ideas from kinetic proofreading, an error-correcting mechanism in biology, to identify new kinetic regimes in non-equilibrium systems. These regimes are defined by the sensitivity of the occupancy of a state of the system to a change in its energy. In biological contexts, higher sensitivity corresponds to stronger discrimination between molecular substrates with different energetics competing in the same reaction. We study this discriminatory ability in systems with discrete states that are connected by a general network of transitions. We find multiple regimes of different discriminatory ability when the energy of a given state of the network is varied. Interestingly, the occupancy of the state can even \textit{increase} with its energy, corresponding to an ``anti-proofreading'' regime. 
The number and properties of such discriminatory regimes are limited by the topology of the network.  Finally, we find that discriminatory regimes can be changed without modifying any ``hard-wired'' structural aspects of the system but rather by simply changing external chemical potentials.
\end{abstract}

\maketitle
Non-equilibrium systems often show varied behavior that depends on many parameters and details of the system. While it may be easy to investigate any particular point in the parameter space of a given system, it is harder to identify qualitative regimes relevant across systems. For example, the occupancy of states at thermal equilibrium is simply given by the Boltzman distribution $\psi_{eq}(E) \sim e^{-\frac{E}{k_{B} T}}$ which depends only on the energy $E$ of a state and the temperature $T$. When driven out of equilibrium, the occupancy of states could in principle depend on all the details of the space of states of the system and the network of paths connecting them. There has been much effort invested \cite{Bustamante05, Seifert08, Min04, Zia11} in understanding non-equilibrium steady states (their occupancy of states, fluxes and entropy production) in order to model the many systems in physics \cite{Schnakenberg:1976wb, Zia06, Zia07, Derrida07, Seifert07,Vaikuntanathan:2013tt} and biology \cite{Hill89, Qian05b, Qian06a, Seifert11,Tu:2008vn,Lan:2012in}  that operate out of equilibrium.

In this paper, we identify steady-state kinetic regimes
that can be exhibited by general systems that are driven
out of equilibrium.  These regimes are characterized by
the change in the occupancy of a state in response to
a change in its energy, with all driving forces held fixed.

We draw on intuition from generalized versions of a biological error correcting mechanism called kinetic proofreading. In biochemical contexts, two molecular substrates, one of them undesirable, might compete to participate in the same enzymatic reaction with different binding energies. Hopfield and Ninio \cite{Hopfield:1974uo,Ninio:1975vv}  proposed a non-equilibrium ``kinetic proofreading'' mechanism which enhances the effect of the binding energy difference on reaction rates. As a result, reactions with the weakly binding substrate are suppressed to a much larger extent than one would expect at equilibrium. Kinetic proofreading has been evoked to explain very low error rates in a host of biochemical reactions despite the similarity of competing substrates; in DNA replication despite the similarity of the four nucleic acids, in protein synthesis despite the similarity of t-RNA molecules \cite{Hopfield:1974uo,Ninio:1975vv, Ehrenberg12, Rodnina04}, in immune response that discriminates between native and foreign proteins \cite{McKeithan95} and in many other biological phenomena \cite{Tu:2008vn,Tsvi04, Qian06a,Qian12}. In all of these cases, proofreading enhances discrimination between competing substrates by making the occupancy of a biochemical state more sensitive to changes in its energy. 

The central proposition of our paper is that non-equilibrium systems like proofreading mechanisms can be designed to have co-existing regimes of varying discrimination. In the context of proofreading, some undesirable substrates can be highly discriminated against while other substrates of even weaker binding energy can actually be promoted. The number and properties of such co-existing discriminatory regimes are constrained by the topology of the network of paths connecting the system's states. Within these constraints, we can design variable discrimination through an appropriate choice of kinetic parameters that might be ``hardwired'', for example, in the structural and dynamical properties of molecules. 

On the other hand, non-equilibrium steady states are always powered by external sources of free energy, such as ATP hydrolysis in the case of proofreading. We find that merely changing the chemical potential of these external sources can change the discrimination characteristics. The ability to tune such characteristics on the fly, without having to change any hardwired properties, raises interesting ways in which natural and synthetic systems can respond to the environment.

\section{One-loop network}
\begin{figure}
\centerline{\includegraphics[scale=0.7]{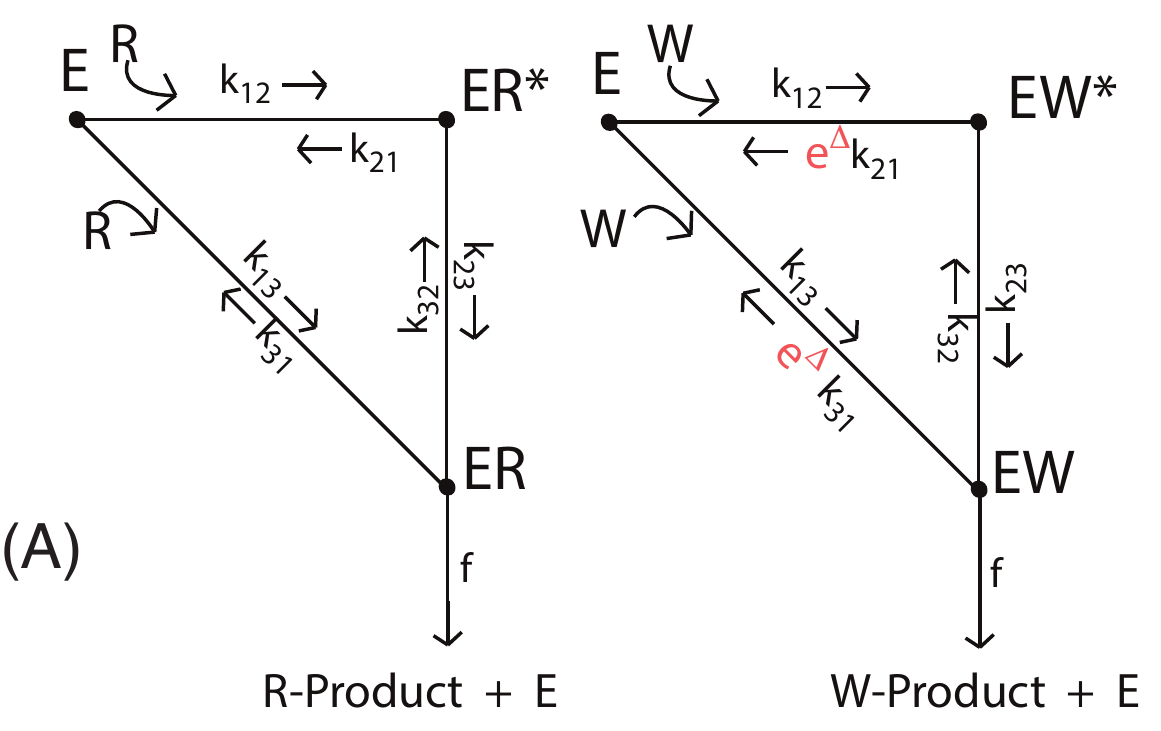}}
{\includegraphics[scale=0.72]{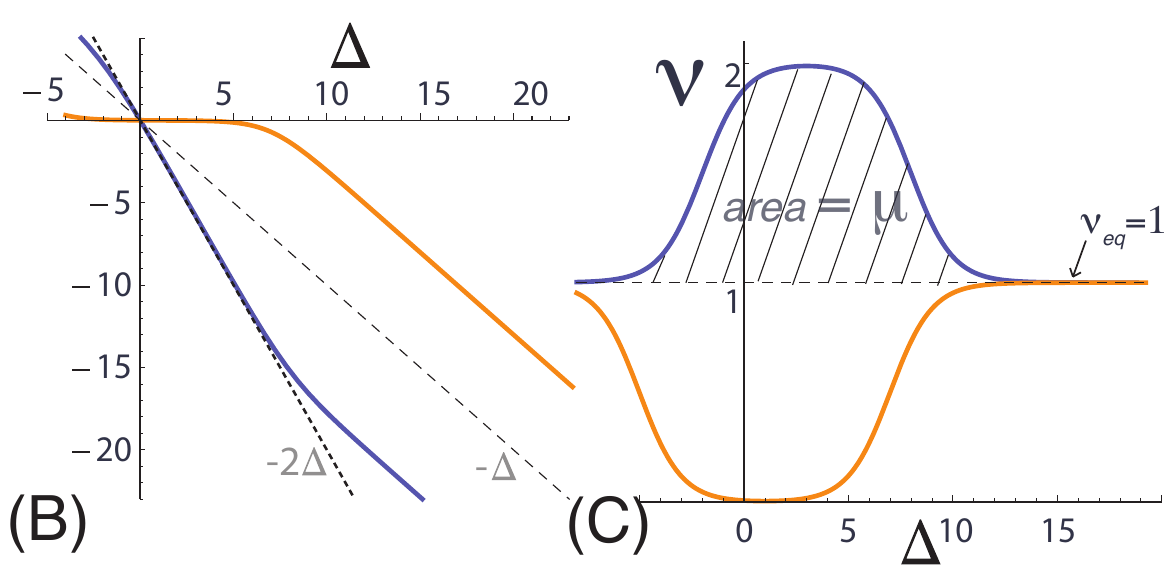}} 
\caption{(A) The kinetic proofreading scheme of Hopfield and Ninio \cite{Hopfield:1974uo,Ninio:1975vv} shows variable discrimination between two substrates $W$ and $R$ that compete to participate in the same reactions with an enzyme $E$ but with binding energies that differ by $\Delta$. 
(B) The ratio $\eta(\Delta)$ of $W$-Product to $R$-Product, with kinetic parameters in the limit proposed by \cite{Hopfield:1974uo,Ninio:1975vv}, is shown in blue. While a substrate $W$ is suppressed to the extent of $e^{-2 \Delta}$ for $\Delta < 8 k_B T$, proofreading is less effective for larger $\Delta$. The orange curve in (B) corresponds an alternative limit of kinetics, distinct from that in \cite{Hopfield:1974uo,Ninio:1975vv}. Instead of correcting errors, the non-equilibrium network promotes the formation of the less stable $W$-Product for $\Delta < 6 k_B T$, with $\eta(\Delta) \sim 1$. (C) The discriminatory index $\nu$ (i.e., slope of $\log(\eta)$) shows distinctive discriminatory regimes for both limits of kinetics. The area between the graph of $\nu$ and the equilibrium value $\nu_{eq} = 1$ is equal to the chemical potential $\mu$ used to drive the network. (The area of the orange curve is interpreted as a negative number.) \label{fig:HopfieldNetwork}}
\end{figure}

\begin{figure}
\centerline{\includegraphics[scale=0.42]{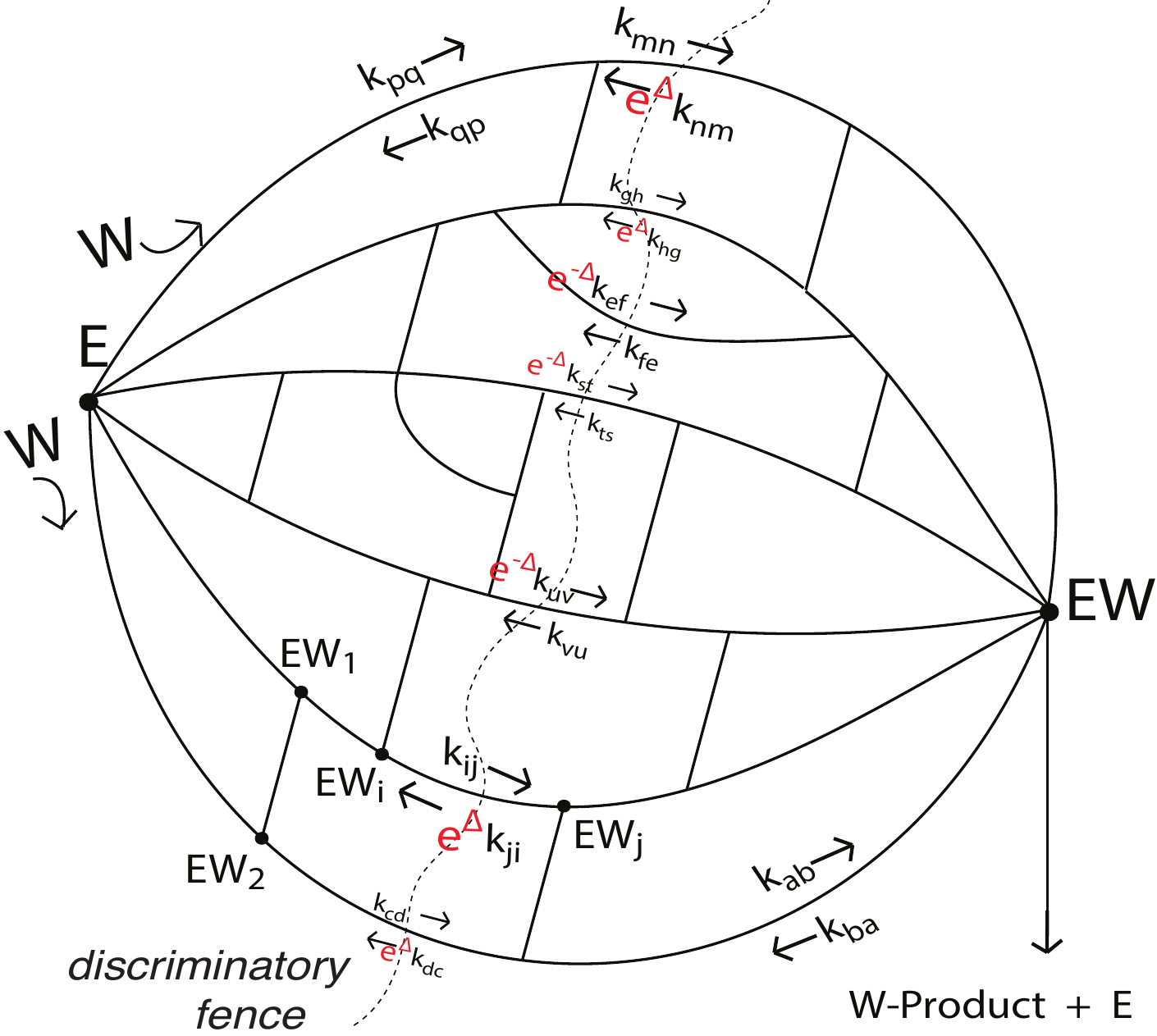}}
\caption{A general proofreading network with multiple paths from $E+W$ to $EW$. We have shown only the kinetics for substrate $W$. The kinetics $k_{ij}$ for $R$ are identical to $W$'s, except along a set of $c=7$ reactions that cross a ``discriminatory fence'' (dotted line) and account for the binding energy difference $\Delta$ between $EW$ and $ER$. Reactions on either side of the fence do not discriminate between $W$ and $R$. When driven out of equilibrium, the occupancy of $EW$ can be made as low as $\eta(\Delta) \sim e^{-7 \Delta}$ or as high as $e^{5 \Delta}$ (relative to $ER$).
\label{fig:GenNetwork}}
\end{figure}

We can illustrate many of our results using the simple 3-state system shown in Fig \ref{fig:HopfieldNetwork}A, with stochastic transitions governed by kinetic rates $k_{ij}$. We will then generalize these results to non-equilibrium networks with many more states and pathways (Fig.~\ref{fig:GenNetwork}). The model shown in Fig.~\ref{fig:HopfieldNetwork}A was introduced by Hopfield and Ninio \cite{Hopfield:1974uo,Ninio:1975vv} as a non-equilibrium solution to the following discrimination problem in biochemistry: Enzyme $E$ reacts with a substrate $R$, forms complexes $ER,ER^*$, and thus processes $R$ into a biologically active product (R-Product). However, a structurally similar competing substrate $W$ might also be able to undergo the same set of reactions with $E$, with less favorable energetics, releasing an undesirable product $W$-Product. 

The steady-state occupancies $\psi_{ER}, \psi_{ER*}, \psi_{E}$ are given by solving the master equation,
\beqa
\partial_t \psi_i = \sum_{j} w^R_{ji} \psi_j - w^R_{ij} \psi_i = 0. 
\eeqa 
where $w^R_{ij}$ are the kinetic constants $k_{ij}$ shown on the network to the left in Fig \ref{fig:HopfieldNetwork}A. Occupancies $\psi_{EW}, \psi_{EW*},\psi_{E}$ of the $W$-network are given by solving the same equation with modified kinetic constants $w^W_{ij}$ shown to the right in Fig \ref{fig:HopfieldNetwork}A. The kinetics of $W$ and $R$ differ because the binding energy of $EW$ is lower than $ER$ by $\Delta$; hence the off-rates $w^W_{21}, w^W_{31}$ are higher than the corresponding rates for $R$ by a factor of $e^{\Delta}$. We do not consider the case of discrimination due to variable activation barriers studied in \cite{Bennett:1979tb,Sartori:2013fv}.

The error rate is the ratio of $W$-Product to $R$-Product formation, given (in the limit of slow formation rate $f$ from the complexes $EW$, $ER$) by \cite{Hopfield:1974uo}, 
\begin{equation}
\eta(\Delta) = \frac{\psi_{EW}}{\psi_{E}}(\Delta) \times \frac{\psi_{E}}{\psi_{ER}}.
\end{equation}
If the system is at equilibrium and not driven, the rates $w^R_{ij},w^W_{ij}$ will both satisfy detailed balance: we can write ${w^S_{ij}}/{w^S_{ji}} = e^{-(V^S_i - V^S_j)} $ where $S = R$ or $W$ and $V^S_i$ is the energy of state $i$. (We have set $\beta = {1}/{k_B T} =1$.) It is easy to see that $\eta(\Delta) = e^{-\Delta}$, the Boltzmann distribution. 

Hopfield and Ninio proposed that discrimination between $R$ and $W$ can be much higher if the system is driven out of equilibrium by coupling some of the reactions to ATP hydrolysis. The rates $w^S_{ij}$, which take ATP coupling into account, will no longer obey detailed balance but can written as,
\beqa
\frac{w^S_{ij}}{w^S_{ji}} = e^{-(V^S_i - V^S_j) + \mu_{ij}}. \label{eqn:brokenDB}
\eeqa
Here $S=R$ or $W$. $\mu_{ij}$ is the external driving potential on the $i \leftrightarrow j$ reaction due to ATP coupling and is conservatively assumed to not distinguish between $W$ and $R$. Hopfield and Ninio identified a particular limit of kinetics $w^R_{ij},w^{W}_{ij}$ in which such driving would result in an error rate $\eta \sim e^{-2 \Delta}$ that is typically significantly lower than the equilibrium error rate $e^{-\Delta}$. This mechanism has been invoked to explain low error rates in biochemical processes like protein synthesis \cite{Johansson:2008tz}.

Physically, the error rate $\eta(\Delta)$ is a measure of how much the occupancy $\psi_{ER}$ of state $ER$ changes when the energy of $ER$ is raised by $\Delta$, with external driving forces $\mu_{ij}$ held fixed. A larger change in occupancy implies higher discriminatory power between substrates in the biochemical reaction. Hence we will find it useful to define a local measure of this discriminatory ability, the discriminatory index,
\beqa
\nu(\Delta) \equiv - \frac{\partial \log \eta(\Delta) }{\partial \Delta}. \label{eqn:defnu}
\eeqa
At equilibrium, $\nu(\Delta) = 1$ while $\nu(\Delta) = 2$ in the kinetic limit uncovered by Hopfield and Ninio.

\textbf{Discriminatory regimes}
The central observation of our paper is that $\eta$ and $\nu$ are not constant but rather functions of $\Delta$. The full plot of $\log \eta(\Delta)$ with kinetic parameters in Hopfield's \cite{Hopfield:1974uo} proposed limit is shown in Fig.~\ref{fig:HopfieldNetwork}B (blue curve). 
The classical result of $\eta \sim e^{-2 \Delta}$ only holds for $\Delta < 8 k_B T$ (in this example) after which $\log \eta(\Delta)$ quickly transitions to a different regime with slope $-1$ (the equilibrium value). Proofreading is less effective for substrates with $\Delta$ in this regime. 
 We say that the network exhibits two co-existing ``discriminatory regimes'', i.e, two distinct regions of $\Delta$ characterized by $\nu \approx 2$ and $\nu \approx 1$ respectively.

\textbf{Malleability of regimes:}
If we choose an alternative set of kinetic constants $k_{ij}$, distinct from the limit in \cite{Hopfield:1974uo,Ninio:1987vi}, we find a qualitatively different set of regimes with $\nu \approx 0$ and $\nu \approx 1$, shown in Fig \ref{fig:HopfieldNetwork}B,C (orange). In this sense, discriminatory regimes are malleable and can be changed through an appropriate choice of kinetic constants. (See SI for the values of $k_{ij}$ for both curves and for a 2-loop network example.)

\textbf{Anti-proofreading:} The kinetic limit corresponding to $\nu < 1$ (Fig.~\ref{fig:HopfieldNetwork}C orange) uses non-equilibrium effects to actually \emph{lower} the discrimination between $W$ and $R$; the weakly binding substrate $W$ with $\Delta < 6 k_B T$ forms as much product as strongly binding $R$ since $\nu \approx 0$ and $\eta(\Delta) \sim 1$. This is an example of an ``anti-proofreading'' regime, in which higher energy states can have higher occupancy.

\textbf{Size of regimes and chemical potentials:}
The size of the discriminatory regimes shown in Fig.~\ref{fig:HopfieldNetwork}C can be quantified by the area between the graph of $\nu(\Delta)$ and the equilibrium value $\nu_{eq}=1$. 
This area is in fact equal to the net chemical potential around the loop $\mu = \mu_{12} + \mu_{23} +\mu_{31}$ that is used to drive the system. Thus higher chemical potentials are needed to enhance discrimination across a wider range of energies. 
 
\textbf{External shaping of regimes:}
The relationship between chemical potential and the shape of $\eta(\Delta)$ and $\nu(\Delta)$ implies that we can modify discriminatory regimes without having to change any kinetic constants  ``hard-wired'' in the structure of the enzyme. For example, we can switch from the kinetics corresponding to the blue curve in Fig.~\ref{fig:HopfieldNetwork}C to the orange curve's by flipping the sign of chemical potentials $\mu_{ij}$ --- i.e., by driving the system in the reverse direction around the reaction loop. 

\section{General networks}

We will now show how the kinetics of a general driven network can give rise to the variable discriminatory regimes discussed above. We will find that topology of the network limits the number, size and index $\nu$ of discriminatory regimes that can be achieved.

A general driven network with multiple interconnected pathways is shown in Fig.~\ref{fig:GenNetwork}. We have only shown the kinetics for substrate $W$, whose binding energy to $E$ is lower than that for $R$ by $\Delta$. This binding energy difference can be reflected in the kinetics in many different ways. The only requirement is that the equilibrium constant computed along any path from start ($E+S$) to finish ($ES$) for $S=R$ and $S=W$ must differ by a factor of $e^\Delta$. For now, we assume that the distinction in kinetics is localized in one reaction along each pathway as shown in Fig.~\ref{fig:GenNetwork}. We connect such reactions by an imaginary line that we call the ``discriminatory fence'' (dotted line in Fig.~\ref{fig:GenNetwork}), which divides the network in two. Reactions on either side of the fence do not discriminate between the two substrates; the binding energy difference $\Delta$ is entirely accounted for by reactions that cross the discriminatory fence. (The fence is a surface for non-planar networks. See SI for a discussion.)

Discrimination in several biological proofreading mechanisms does appear to be localized in select reactions \cite{AltanBonnet:2005bz,Johansson:2008tz,BarZiv:2002jd}. We relax this simplifying assumption later and find that a de-localized fence lowers the discriminatory ability of the network.

\subsection{Limit of highest discrimination}
To gain intuition about general networks, we begin with the kinetic limit that leads to the highest possible discriminatory index $\nu > 0$ (strongest proofreading) in any network. This kinetic limit generalizes several particular models of biological error correction \cite{Tu:2008vn,McKeithan:1995wq,AltanBonnet:2005bz,Savageau:1981tz, BarZiv:2002jd} and is defined by three properties:

\begin{figure}
\includegraphics[scale=0.4]{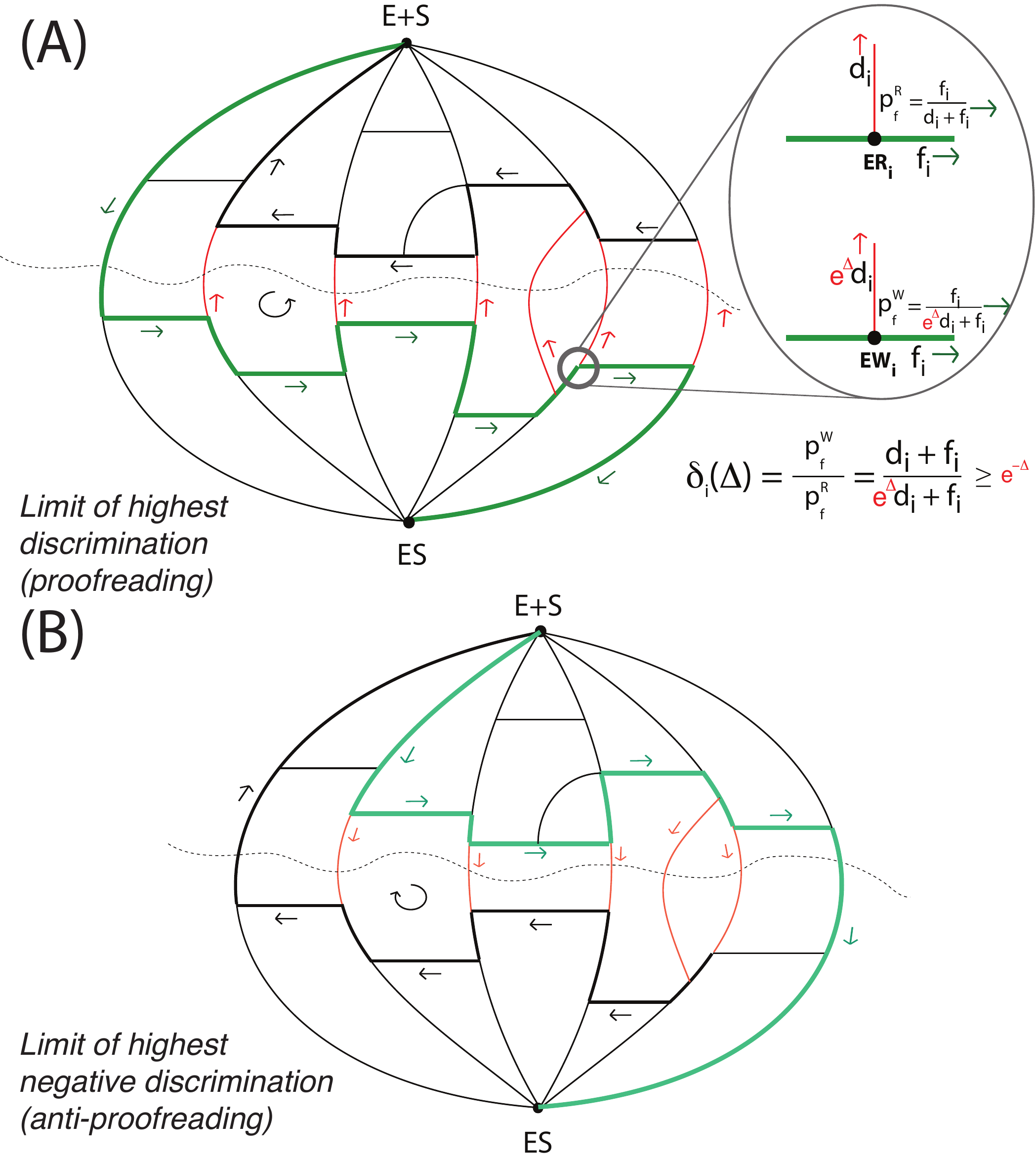}
\caption{General limit of kinetics for highest (A) discrimination $\nu > 0$ and (B) negative discrimination $\nu < 0$ (anti-proofreading). In both cases, the system can reach $ES$ only by traveling down a dominant path (green) that parallels the discriminatory fence (dotted line). In this way, the system is exposed to multiple discard pathways (shown in red) that cross the discriminatory fence (dotted line) and abort the reaction by taking the system back to the initial state $E+S$ through the bold black path (here, $S=R$ or $W$). (A) As shown in the inset, the probability $p_f^S$ of not taking a red discard path across the fence is lower for $S=W$ than $S=R$ by a factor $\delta_i(\Delta)$. With $c$ such discard paths along the dominant path, $W$ is exponentially less likely to reach $ES$ than $R$, resulting in $\eta(\Delta) \sim e^{-c \Delta}$. In (B), the dominant path parallels the fence on the reactants $E+S$ side of the network. The system can now take the (red) discard paths from the \emph{reactants side to the product side}. Unlike in (A), we allow for the system to possibly bounce back to the dominant path after a discard. As shown in the SI, $W$ is more likely to bounce back and proceed further along the dominant (green) path by a factor $\delta_i(\Delta) > 1$. Hence the weakly binding $S=W$ substrate is exponentially more likely to reach $ES$.\label{fig:HighLowErrorPath}}
\end{figure}

(i) The kinetics pick out a dominant path, shown in green in Fig \ref{fig:HighLowErrorPath}A, from state $E+S$ to the final state $ES$; the system can reach $ES$ only by traversing this path. 

(ii) The dominant path must parallel the discriminatory fence on the products (i.e., $ES$) side of the network before reaching $ES$. As a result, as the system travels down the dominant path towards $ES$, it is constantly exposed to pathways (shown in red) that cross the fence back to the reactants ($E+S$) side.

(iii) The kinetics need to ensure that if the system crosses the fence to the reactants side along a (red) discard pathway, the system moves back towards $E+S$ along a path like that shown in bold.  Hence we refer to the (red) pathways as discard pathways since they stop progress along the dominant path.

Hence a typical trajectory in Fig.~\ref{fig:HighLowErrorPath}A involves the system attempting to traverse the green path to the final state $ES$ but being frequently discarded through the red fence-crossing reactions. The kinetics described thus far apply to both $S=R$ and $S=W$ substrates. The only distinction is that $W$ is more likely take each discard pathway across the discriminatory fence. To see this, note that the probability of being discarded is determined by the kinetic constants $d_i$ for the discard pathway and $f_i$ for the forward direction on the dominant pathway (see inset in Fig.~\ref{fig:HighLowErrorPath}A). Since $d_i$ is higher for $W$ than for $R$ by a factor $e^\Delta$, $W$ is less likely to proceed forward along the dominant path by a factor $\delta_i$ shown in Fig.~\ref{fig:HighLowErrorPath}A.

In the limit of high discards $ d_i \gg f_i$, the ratio of the forward probabilities $\delta_i$ is $e^{-\Delta}$. If there are $c$ such discard paths branching off the dominant path, the net probability of reaching $ES$ is lower for $W$ than for $R$ by a factor $\eta(\Delta) \sim e^{-c \Delta}$, giving the highest possible discriminatory index $\nu = c$.

We note that much time is wasted in this limit since both $R$ and $W$ are frequently discarded along the dominant path and reaction completion is an exponentially unlikely event. Energy is also wasted since the system preferentially executes counterclockwise trajectories in the network shown in Fig.~\ref{fig:HighLowErrorPath}; such preferential cycling breaks detailed balance and consumes free energy from an external source (such as ATP hydrolysis) during each cycle. An alternative kinetic regime which trades a small increase in error for a large saving in time and energy was found in \cite{Murugan:2012dz}.

\textbf{Anti-Proofreading:} 
If the dominant path parallels the discriminatory fence on the \emph{reactants side} ($E+S$) side instead (Fig.~\ref{fig:HighLowErrorPath}B), we find a counterintuitive ``anti-proofreading'' regime with $\nu < 0$ and $\eta \sim e^{+(c-2)\Delta}$. The discard pathways now cross the fence from the reactants to product side. Unlike in the proofreading case, we assume that there is a high chance of reversing the discard reaction immediately. As a result, the system bounces back and forth repeatedly along a discard path before either proceeding forward along the dominant path or going back to the initial state along the bold black path in Fig.~\ref{fig:HighLowErrorPath}B. As shown in the SI, substrate $W$ is more likely to bounce back across the fence and resume progress along the dominant path by a factor $\delta_i(\Delta) > 1$. Hence the weakly-binding substrate $W$ reaching the final state $EW$ more often than $R$ reaches $ER$. In other words, the occupancy of the state $ES$ increases as its energy is raised.

\subsection{Sculpting multiple discriminatory regimes}
The kinetic pathways for highest positive and negative discrimination suggest how variable discriminatory regimes arise for a general network. As shown in the Fig.~\ref{fig:HighLowErrorPath} inset for the case of positive discrimination, each $\delta_i(\Delta) = (d_i + f_i)/ (d_i e^\Delta + f_i)$ provides varying discrimination that depends on $\Delta$ and the discard $d_i$ and forward $f_i$ kinetics at that point. In the limits described in Fig.~\ref{fig:HighLowErrorPath}, the occupancy ratio $\eta(\Delta)$ is (approximately) a product of $\delta_i$ along the dominant path;
\begin{equation}
\eta(\Delta) \sim \delta_1(\Delta)  \delta_2(\Delta) \ldots \delta_c(\Delta).
\label{eqn:etadelta}
\end{equation} 

\begin{figure}
\centerline{\includegraphics[scale=0.25]{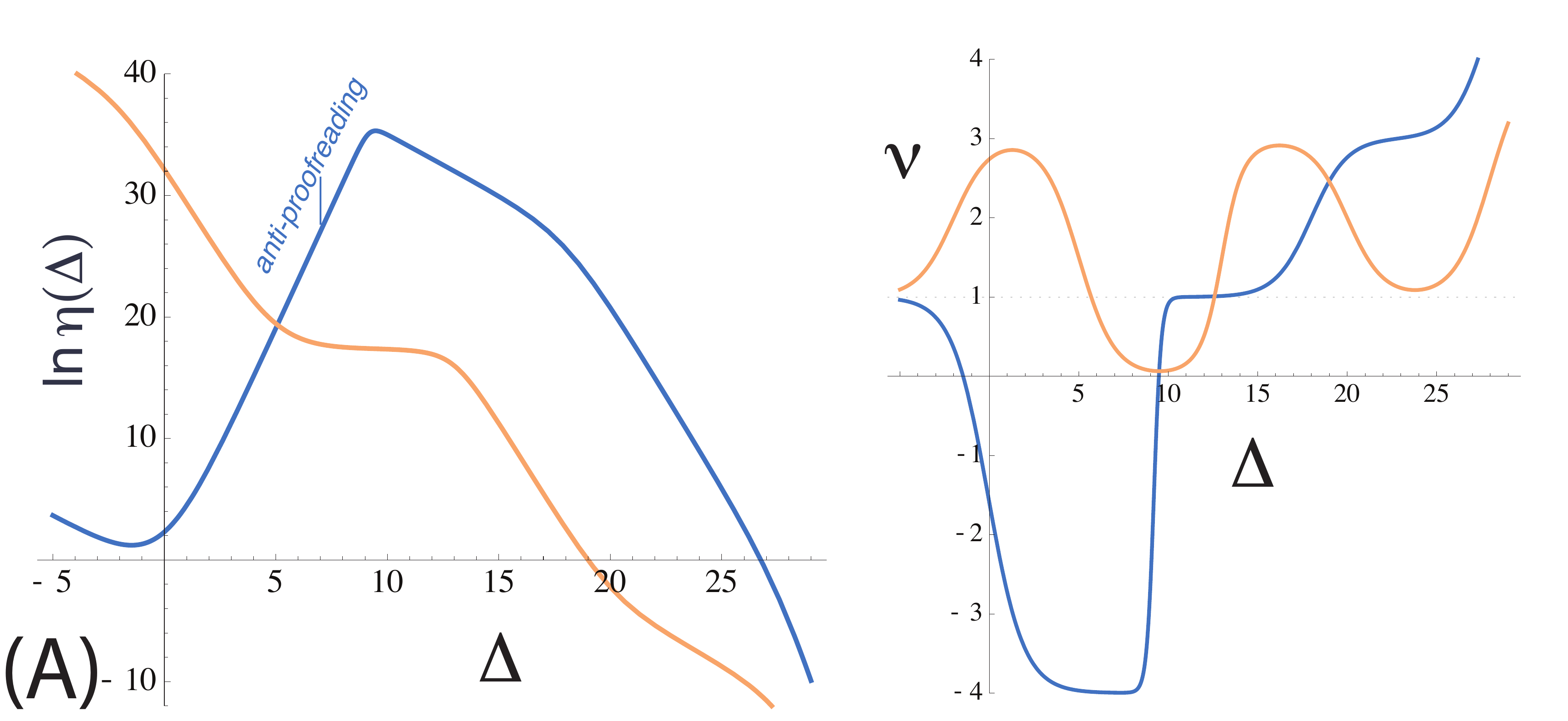}}
\centerline{\includegraphics[scale=0.5]{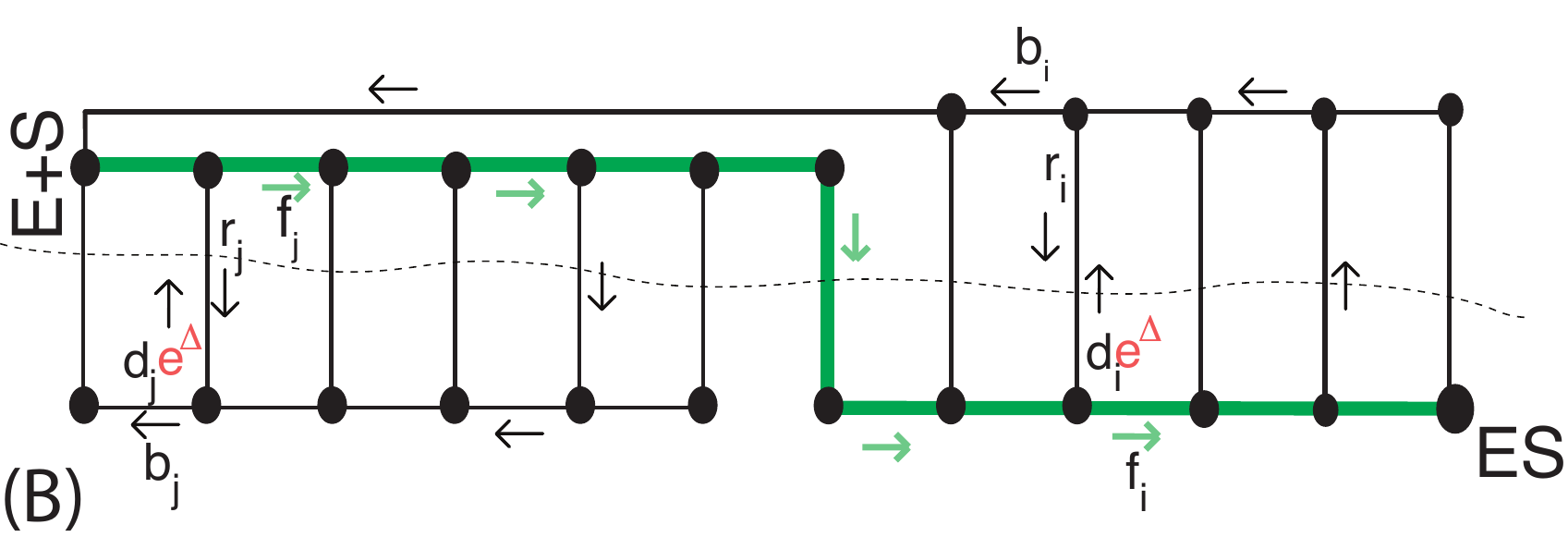}}
\caption{ (A) We can sculpt $\eta(\Delta),\nu(\Delta)$ shown, with multiple discriminatory regimes, by combining the two limits shown in Fig.~\ref{fig:HighLowErrorPath}. We use a ladder-like network shown in (B) and a dominant path that switches midway from paralleling the fence on the reactants side (like in Fig.~\ref{fig:HighLowErrorPath}B) to the products side (like in Fig.~\ref{fig:HighLowErrorPath}A). We position the regimes as shown (blue and orange plots) through two different choices of kinetic constants $d_i,f_i,r_i,b_i$. These constants determine $\delta_i(\Delta)$ (see Fig.~\ref{fig:HighLowErrorPath} and Eq.~\ref{eqn:etadelta}) along the dominant path. (See SI for numerical values of $d_i,f_i$ and more on $\delta_i(\Delta)$. We added a constant to the orange curve in (A) to fit both curves in the same plot.) \label{fig:nLoopIdx}}
\end{figure}

As a result, we can design a large network to show multiple discriminatory regimes like that shown in Fig.~\ref{fig:nLoopIdx}A through a choice of dominant path and a choice of $\delta_i$s along the path. In Fig.~\ref{fig:nLoopIdx}B, the dominant path switches midway from paralleling the discriminatory fence on the reactants side (like in Fig.~\ref{fig:HighLowErrorPath}B) to the products side (like in Fig.~\ref{fig:HighLowErrorPath}A), allowing a more general product of $\delta_i$. (See SI for details on $\delta_i$ on anti-proofreading dominant paths.)

More generally, $\eta$ is a ratio of polynomials ${p(e^\Delta)}/{q(e^\Delta)}$ in $e^\Delta$. A discriminatory regime with index $\nu=j-i$ corresponds to a finite range of $\Delta$ over which $\eta(\Delta)$ can be approximated by ${a_i e^{i \Delta}}/{b_j e^{j \Delta}}$ where $a_i e^{i \Delta}$ and $b_j e^{j \Delta}$ are particular monomials in $p$ and $q$ respectively. (See SI for a rigorous derivation of these results using Schnakenberg's network theory \cite{Schnakenberg:1976wb}.)

\subsection{Network topology constrains regimes}
With $\eta(e^\Delta) = {p(e^\Delta)}/{q(e^\Delta)}$, we find that the degrees of the polynomials $p,q$ are bound by $c$, the number of intersections between the discriminatory fence and the network. As a result, topological properties of the proofreading reaction network constrain the shape of $\eta(\Delta)$ and index $\nu(\Delta)$:

(i) The largest positive and negative values of $\nu$ are limited by $c$: $ -(c-2) < \nu < c$.  

(ii) The number of distinct discriminatory regimes - determined by the number of monomials in $p,q$ - is also bounded by $c$. 

(iii) The extremal values attained by $\nu$ over all $\Delta$ are related by $\nu_{max} - \nu_{min} \leq c$.

(iv) The intersection number $c$ cannot exceed the number of linearly independent pathways. Equivalently, in terms of the number of linearly independent loops $n$ in the network, $ c \leq n+1$.

(v) This sets an absolute limit on the discrimination between two substrates whose binding energies differ by $\Delta$ by a proofreading network with $n$ loops:
\begin{equation}
e^{(n-1) \Delta} \geq \eta(\Delta) \geq e^{-(n+1) \Delta}.
\end{equation}
If the discriminatory fence intersects only $c < n+1$ reactions, then $e^{ (c-2) \Delta} \geq \psi(\Delta) \geq e^{ -c \Delta}$. For the network shown in Fig \ref{fig:GenNetwork}, $c = 7$ while $n = 18$. 

(vi) We find that the minimal chemical potentials required for a discriminatory index $\nu(\Delta)$ are given  by the areas of the $n$ discriminatory regimes in the graph of $\nu - 1$. When $\nu < 1$, the area and chemical potential required are negative, indicating that the corresponding loop must be driven in the opposite direction. We leave a study of the precise mapping between the $n$ linearly independent chemical potentials $\mu_l$ in a network with $n$ loops and the $n$ resulting regimes to future work.

\subsection{Localized discrimination fence and von Neumann's error correcting scheme}

\begin{figure}
\centerline{\includegraphics[scale=0.6]{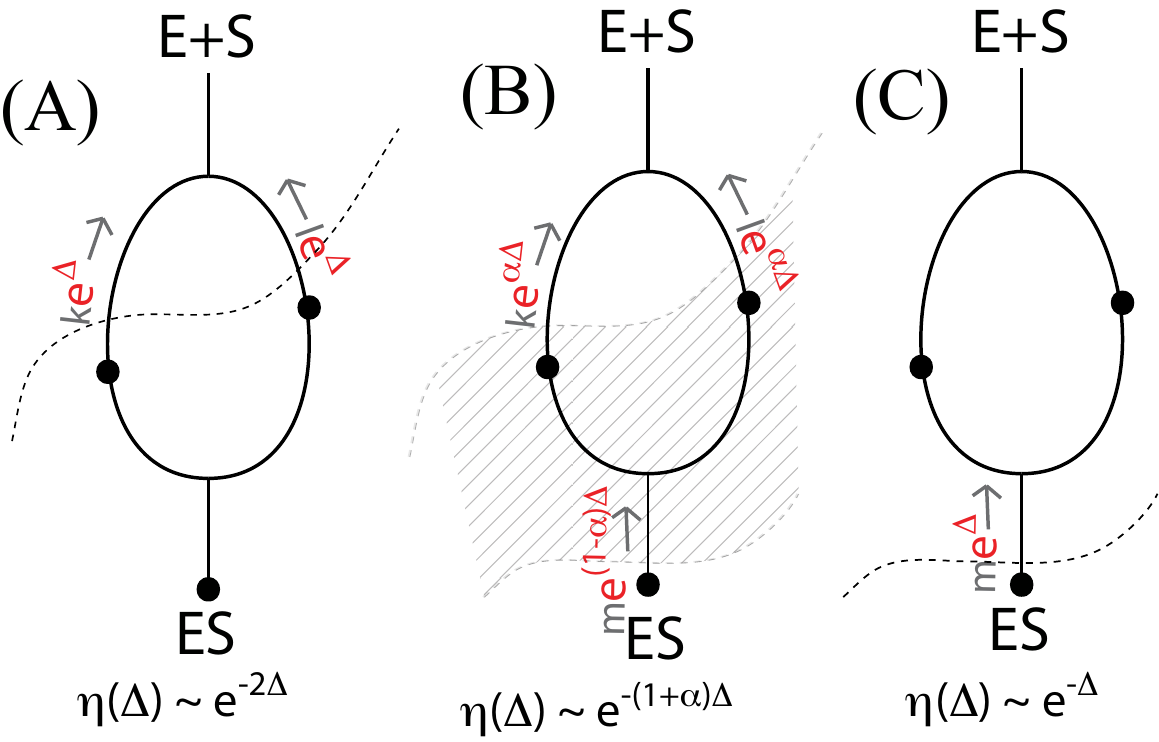}}
\caption{ (A) Discrimination is highest when the energy difference $\Delta$ is localized to a discriminatory fence that cuts through as many reactions as possible. In (B) and (C), a fraction $\alpha$ or all of energy difference $\Delta$ is in reactions common to both pathways ($0\leq\alpha\leq 1$). Proofreading acts by multiplying the effect of discrimination in \emph{parallel and independent} reactions but cannot enhance discriminatory reactions common to both pathways.
\label{fig:IndepVsDep}}
\end{figure}

We have assumed that the binding energy difference $\Delta$ is localized in a set of reactions, one along each pathway from $E+S$ to $ES$, that define the sharp discriminatory fence. Consider the alternatives shown in Fig \ref{fig:IndepVsDep} where the energy $\Delta$ is (B) spread over multiple reactions along each pathway or (C) where the energy difference $\Delta$ is localized in a reaction common to both pathways. Proofreading can only combine \textit{independent} discriminatory power in parallel pathways that form loops. Hence while the network in Fig.~\ref{fig:IndepVsDep}(A) achieves discrimination of $e^{-2 \Delta}$, (B) and (C) can only achieve lower discrimination of $e^{- (1-\alpha) \Delta} \times e^{-2 \alpha \Delta} = e^{-(1+\alpha) \Delta}$  and $e^{-\Delta}$ respectively. (In (B), we assume $0 \leq \alpha \leq 1$ so that no individual reaction has discriminatory power greater than $e^{-\Delta}$.) 

We can think of proofreading as a biochemical implementation of von Neumann's \cite{vonNeumann56} reliable machine made of redundant unreliable components. von Neumann constructed logical machines that failed only when all of the individual error-prone components failed. Thus the machine itself has a lower error rate than the individual components, provided the components have independent probabilities of failure. In proofreading, each pathway can complete the entire reaction in isolation with an error rate of $e^{-\Delta}$ but the network as a whole can have a lower error rate $\eta(\Delta) \ll e^{-\Delta}$. In this context, the two pathways of Fig.~\ref{fig:IndepVsDep}(A) are like von Neumann components with fully independent error rates while the error rates of the pathways in (B) and (C) are not independent and cannot be combined as effectively.

\section{Discussion}

By generalizing kinetic proofreading, we have uncovered new kinetic regimes in systems driven out of equilibrium, characterized by how much the occupancy of a state changes due to a change in its energy. The occupancy can fall with energy much faster than at equilibrium (conventional proofreading) but we also found anti-proofreading regimes where the occupancy of a state increases with increasing energy. We found bounds on the number, size and discriminatory ability of regimes in terms of topological properties of the network of transitions between states.

We were able to identify kinetic limits associated with discriminatory regimes in terms of a dominant path that parallels the discriminatory fence (the collection of reactions whose kinetics account for the change in energy of a state). We need such an understanding of kinetics that is not tied to a particular network in order to be able to incorporate non-equilibrium driving into synthetic systems to achieve particular goals. Such a general picture is also useful in studying proofreading properties of biochemical circuits, which are often large, complex and designed to perform several functions besides proofreading.

Biologically, having proofreading ability vary with binding energy can be useful when a biochemical reaction is presented with a family of competing substrates. The same proofreading scheme might be able to enhance discrimination between two given substrates while in fact suppressing differences between another pair, depending on their binding energies.

We emphasize that our results apply to networks that represent any driven system, including systems unrelated to biological kinetic proofreading. For example, the states of the network could represent different stages of assembly of microtubules \cite{Murugan:2012dz} or synthetic self-assembled structures \cite{Wei:2012ip,King:2012gc}. Our results then might point the way to incorporating driving forces into the assembly pathways in synthetic systems, reducing errors in the final structure.

An intriguing feature of the schemes discussed here is the possibility of changing the discriminatory regimes by changing only the chemical potentials $\mu_i$, without having to change any hard-wired structural kinetics. Such an ability to change discriminatory properties on the fly could be useful in both natural and synthetic systems; the network can switch between suppressing the reaction with a substrate and promoting it, depending on the environment or other considerations. 

\begin{acknowledgments}
We would like to thank John J.Hopfield and Luca Peliti for stimulating discussions. We are also grateful to Luca Peliti for detailed comments on a draft of this paper. 
\end{acknowledgments}


\end{document}